# ASSESSMENT OF GOES IMAGER MICROBURST PRODUCT OVER THE SOUTHWESTERN UNITED STATES

Kenneth L. Pryor
Center for Satellite Applications and Research (NOAA/NESDIS)
Camp Springs, MD

## 1. INTRODUCTION

Multispectral GOES imager (Pryor 2009) and Moderate Resolution Imaging Spectroradiometer (MODIS) products have been developed and experimentally implemented to assess downburst potential over the western United States. The availability of the split-window channel (12µm) in both MODIS and the GOES-11 imager allows for the inference of boundary layer moisture content. Wakimoto (1985), based on his study of microbursts that occurred during the Joint Airport Weather Studies (JAWS) project, noted favorable environmental conditions over the western United States: (1) intense solar heating of the surface and a resulting superadiabatic surface layer; (2) a deep, dry-adiabatic convective boundary layer (Sorbjan 1989) that extends upward to near the 500mb level; (3) a well-mixed moisture profile with a large relative humidity gradient between the mid-troposphere and the surface. The GOES-West (GOES-11) imager microburst algorithm employs brightness temperature differences (BTD) between band 3 (upper level water vapor, 6.7µm), band 4 (longwave infrared window, 10.7µm), and split window band 5 (12µm). Band 3 is intended to indicate mid to upper-level moisture content and advection while band 5 indicates low-level moisture content. Soden and Bretherton (1996) (SB96), in their study of the relationship of water vapor radiance and layer-average relative humidity, found a strong negative correlation between 6.7µm brightness temperature ($T_b$) and layer-averaged relative humidity (RH) between the 200 and 500-mb levels. Thus, in the middle to upper troposphere, decreases in $T_b$ are associated with increases in RH as illustrated in Figure 4 of SB96. It follows that large BTD between bands 3 and 5 imply a large relative humidity gradient between the mid-troposphere and the surface, a condition favorable for strong convective downdraft generation due to evaporational cooling in the sub-cloud layer. In addition, small BTD between bands 4 and 5 indicate a relatively dry surface layer with solar heating in progress. Thus the GOES imager microburst risk (MBR) product is based on the following algorithm in which the output brightness temperature difference (BTD) is proportional to microburst potential:

$$MBR(BTD) = \{T_5 - T_3\} - \{T_4 - T_5\} \quad (1)$$

Where the parameter $T_n$ represents the brightness temperature observed in a particular imager band. The relationship between BTD and microburst risk in the product image is based on the following assumptions: (1) A deep, well-mixed convective boundary layer exists in the region of interest; (2) moisture for convective storm development is based in the mid-troposphere and is advected over the region of interest; and (3) the mid- to upper- tropospheric layer of moisture is vertically extensive and would yield precipitation if provided a sufficient forcing mechanism. Similarly, the MODIS microburst algorithm incorporates BTDs between bands 27 (6.535 - 6.895 μm), 31 (10.780 - 11.280 μm) and 32 (11.770 - 12.270 μm). MODIS data is desirable due to its high spatial resolution (1 km). The imager microburst risk product is predictive linear model developed in the manner exemplified in Caracena and Flueck (1988). The microburst product consists of a set of predictor variables that generates output of expected microburst risk. This paper provides further assessment of the GOES imager microburst product, case studies demonstrating effective operational use of the microburst products, and validation results for the 2008 convective season over Arizona.

## 2. METHODOLOGY

The objective of this validation effort was to qualitatively and quantitatively assess the performance of the GOES imager-derived microburst product by employing classical statistical analysis of real-time data. Accordingly, this effort entailed a study of downburst events over central Arizona during the 2008 convective season that was executed in a manner that emulates historic field projects such as the 1982 Joint Airport Weather Studies (JAWS) (Wakimoto 1985). GOES-11 image data was collected for pre-convective environments associated with three downburst events that occurred within the Arizona Automated Local Evaluation in Real Time (ALERT) domain between July and September 2008. In addition, 1 km resolution MODIS data, specifically brightness temperature in bands 27, 31, and 32, was collected for a downburst event that occurred during August 2008. The Flood Control District of Maricopa County (FCDMC) operates a 24-hour rain, stream and weather gage network which provides "real-time" information pertaining to rainfall, floods and weather conditions in Maricopa County and surrounding counties. This network operates in the National Weather Service ALERT format and is commonly referred to as an ALERT system. The ALERT system uses automatic telemetry gages for data collection that transmit their information to the FCDMC.

Wakimoto (1985) and Atkins and Wakimoto (1991) discussed the effectiveness of using mesonetwork surface observations and radar reflectivity data in the verification of the occurrence of downbursts. Well-defined peaks in wind speed (Wakimoto 1985; Atkins and Wakimoto 1991) were effective indicators of downburst occurrence. It was found that derived product imagery generated one to four hours prior to downburst events provided an optimal characterization of the pre-convective thermodynamic environment over Arizona ALERT domain.

During the 2008 convective season, product images were generated by Man computer Interactive Data Access System (McIDAS)-X where GOES imager data was read and processed, brightness temperature differences were calculated, and risk values overlain on GOES imagery. The image data consisted of derived brightness temperatures from infrared bands 3, 4, and 5, obtained from the Comprehensive Large Array-data Stewardship System (CLASS) for archived data and the McIDAS Abstract

Data Distribution Environment (ADDE) server for real-time data. In addition, microburst algorithm output was also visualized by McIDAS-V software (version 1.0beta1). McIDAS-V was used to visualize output area files that were generated by McIDAS-X from archived data as well as MODIS data in HDF format provided by a NASA website. A contrast stretch and built-in color enhancement were applied to the output images to highlight regions of elevated microburst risk. Visualizing algorithm output in McIDAS-V allowed for cursor interrogation of output brightness temperature and more precise recording of BTD values associated with observed downburst events. This methodology should serve to increase the statistical significance of the relationship between output BTD and microburst wind gust magnitude.

Downburst wind gusts, as recorded by ALERT stations, were measured at a height of 10 to 12 feet above ground level. Archived ALERT mesonet observations are available via the Flood Control District of Maricopa County. Based on the previously derived statistical relationship between output BTD and observed downburst wind gust speed (Pryor 2009), a color-enhanced risk image was generated that indicates increasing microburst potential as a progression from yellow to red shading. For each microburst event, product images were compared to radar reflectivity imagery and surface observations of convective wind gusts. Next Generation Radar (NEXRAD) base reflectivity imagery (levels II and III) from National Climatic Data Center (NCDC) was utilized to verify that observed wind gusts were associated with downbursts and not associated with other types of convective wind phenomena (i.e. gust fronts). Another application of the NEXRAD imagery was to infer microscale physical properties of downburst-producing convective storms. Particular radar reflectivity signatures, such as the rear-inflow notch (RIN)(Przybylinski 1995) and the spearhead echo (Fujita and Byers 1977), were effective indicators of the occurrence of downbursts.

Covariance between the variables of interest, microburst risk (expressed as output brightness temperature) and surface downburst wind gust speed, was analyzed to assess the performance of the imager microburst algorithm. A very effective means to quantify the functional relationship between microburst index algorithm output and downburst wind gust strength at the surface was to calculate correlation between these variables. Thus, correlation between GOES imager microburst risk and observed surface wind gust velocities for the selected events were computed to assess the significance of this functional relationship. Statistical significance testing was conducted, in the manner described in Pryor and Ellrod (2004), to determine the confidence level of correlations between observed downburst wind gust magnitude and microburst risk values. The derived confidence level is intended to quantify the robustness of the correlation between microburst risk values and wind gust magnitude.

## 3. CASE STUDIES

The first documented microburst in the ALERT domain during the 2008 convective season occurred at Magma FRS station in Pinal County during the evening of 19 July 2008 as shown in

Figure 1, the GOES-11 imager derived microburst risk product at 0100 UTC 20 July 2008 with overlying radar reflectivity from Phoenix, Arizona NEXRAD (IWA) at 0343 UTC. During the evening of 19 July, a cluster of convective storms developed over the mountains of southeastern Arizona and then tracked northeastward toward the greater Phoenix area. A convective storm along the leading edge of the cluster, apparent in Figure 1 over Pinal County, 45 miles southeast of Phoenix, produced a downburst with a measured wind gust of 39 knots at Magma FRS ALERT station at 0333 UTC. Near the time of downburst occurrence, the storm appeared as a spearhead echo in radar imagery. This downburst occurred in close proximity to a local maximum in risk (probability), as indicated by the orange shading immediately downstream of the convective storm. Figure 2, a sounding profile derived from Rapid Update Cycle (RUC) model analysis data at 0100 UTC 20 July 2008 over Magma FRS station, displays significant mid-level moisture overlying a dry surface layer that resulted from several hours of strong surface heating and resultant mixing. Apparent in the sounding profile is a large humidity gradient between the mid-troposphere and the surface that fostered strong convective downdraft generation due to evaporational cooling as precipitation descended in the sub-cloud layer. This sounding profile can be considered prototypical for all three downburst events documented in this paper. The ambient environments of all three events were characterized by a well-mixed, convective boundary layer as exemplified by the "inverted-V" sounding profile in Figure 2.

During the afternoon of 9 August 2008, convective storm activity developed over the mountains of central Arizona. An isolated storm along the southeastern periphery of the area of convective storms produced a strong downburst that was observed at Horseshoe Lake ALERT station. Apparent in Figure 3, MODIS and GOES microburst product images at 2035 UTC and 2130 UTC, respectively, with overlying radar reflectivity from Phoenix, Arizona NEXRAD (IWA) at 2238 UTC is a convective storm over northern Maricopa County that produced the downburst at Horseshoe Lake at 2238 UTC. The strong downburst, with a wind gust of 46 knots, occurred in close proximity to a local maximum in risk, as indicated by the orange shading immediately downstream of the convective storm, especially apparent in the MODIS image with 1 km resolution. The wind histogram shown in Figure 4 at Horseshoe Lake ALERT station represents downburst occurrence as a sharp peak in wind gust speed near 1540 LST 9 August. This information, in conjunction with high radar reflectivity (>55 dBZ) associated with the parent convective storm, confirmed that this wind event was clearly associated with a downburst.

Finally, during the afternoon of 11 September 2008, an area of convective storms developed over western Arizona that tracked eastward toward the Phoenix area. A convective storm along the leading edge of the area produced a strong downburst at Gila Bend ALERT station during the afternoon. Figure 5, the GOES-11 imager derived microburst risk product at 2000 UTC 11 September 2008 with overlying radar reflectivity from Phoenix, Arizona NEXRAD (IWA) at 0022 UTC 12 September displayed a convective storm over southwestern

Maricopa County. The storm produced a downburst with a 42 knot wind gust at Gila Bend at 0025 UTC and occurred in close proximity to a local maximum in risk, as indicated by the orange shading immediately downstream of the convective storm. In a similar manner to the Pinal County storm, radar imagery indicated a spearhead echo associated with the downburst observed at Gila Bend.

Examples of the GOES-11 microburst product images available on the [GOES microburst product website](#) are displayed in Figure 6. Both images display convective storm activity developing over central and western Arizona during the afternoon that would produce downbursts during the following one to four hours. The product image at 2130 UTC 9 August 2008 (top) displays a favorable microburst environment near Horseshoe Lake ALERT station (AHL), where a strong convective wind gust would occur about one hour later at 2238 UTC. In a similar manner, the product image at 2000 UTC 11 September 2008 (bottom), indicates a moderate to high risk of microbursts near Gila Bend (AGL), where a downburst would be observed about four hours later at 0025 UTC 12 September.

## 4. STATISTICAL ANALYSIS AND DISCUSSION

Analysis of covariance between the variables of interest, microburst risk (expressed as output brightness temperature) and surface downburst wind gust speed, provided favorable results for the imager microburst product. Favorable results include a correlation of .50 and a small mean difference (2) between microburst risk values and wind gust speed (in knots) for a dataset of three microburst events that occurred during the 2008 convective season. In addition, statistical significance testing revealed a high (84%) confidence level that the correlation did represent a physical relationship between risk values and downburst magnitude and was not an artifact of the sampling process. It is promising that such a small sample size indicated a significant correlation between GOES-11 microburst algorithm output and observed downburst wind gust speeds.

Documentation of microbursts in Table 1 in the Arizona ALERT domain revealed that microburst activity occurred primarily during the afternoon and early evening. This preference for afternoon and evening microburst activity underscores the importance of solar heating of the boundary layer in the process of convective downdraft generation over central Arizona. The well-mixed moisture profile, best characterized as an "inverted-V", and relative humidity gradient that results from diurnal heating fostered a favorable environment for microbursts that would occur due to the evaporation of precipitation in the dry sub-cloud layer. RUC sounding profiles and high radar reflectivity (> 55 dBZ) provided evidence that significant mid-level moisture promoted precipitation loading as an initiating mechanism for downbursts. The combination of precipitation loading and the presence of a relatively deep and dry convective boundary layer favored a microburst environment that was effectively captured by the GOES imager-derived microburst product. As noted in Pryor (2009), the imager microburst algorithm was originally intended for use in dry microburst environments. However, analysis of three downburst events over central Arizona revealed that the imager microburst algorithms are also effective in identifying favorable hybrid microburst environments. The stronger signal for microburst potential during the August 2008 event was especially apparent in the MODIS image as darker orange shading when compared to the corresponding GOES-11 image. The stronger signal most likely results from the higher spatial resolution and reduced noise associated with the MODIS instrument. Thus, derivation of an algorithm that incorporates GOES-11 bands 3, 4, and 5 and MODIS bands 27, 31, and 32 appears to be effective in indicating a favorable thermodynamic environment for microbursts over central Arizona as well as other regions in the intermountain western U.S.

## 5. SUMMARY AND CONCLUSIONS

As documented in Pryor (2009) and in this paper, and proven by statistical analysis, the multispectral GOES imager product has been found to be effective in indicating the potential for dry and hybrid microbursts. Future development effort will entail the validation of the product over the Arizona ALERT domain during the 2009 convective season following the methodology as outlined in this paper. It is expected that such a procedure should yield a statistically significant sample size, from which product performance should be effectively evaluated using classical statistical analysis. This product provides a higher spatial (4 km) and temporal (30 minutes) resolution than is currently offered by the GOES sounder microburst products and thus, should provide useful information to supplement the sounder products in the convective storm nowcasting process.

**2008 Arizona ALERT Correlation:**

| | | | | | |
|---|---|---|---|---|---|
| GOES to measured wind: | 0.50 | | Mean MBR: | | 45 |
| No. of events: | 3 | | Mean Wind Speed: | | 43 |

| Date | Time (UTC) | Measured Wind Speed kt | Location | GOES-11 MBR | Radar Reflectivity dBZ |
|---|---|---|---|---|---|
| **20-Jul-08** | 0332 | 40 | Magma FRS | 43 | 55 |
| **9-Aug-08** | 2239 | 46 | Horseshoe Lake | 45 | 55 |
| **12-Sep-08** | 0025 | 43 | Gila Bend | 47 | 55 |

Table 1. Statistical data produced by Convective Season 2008 validation effort.

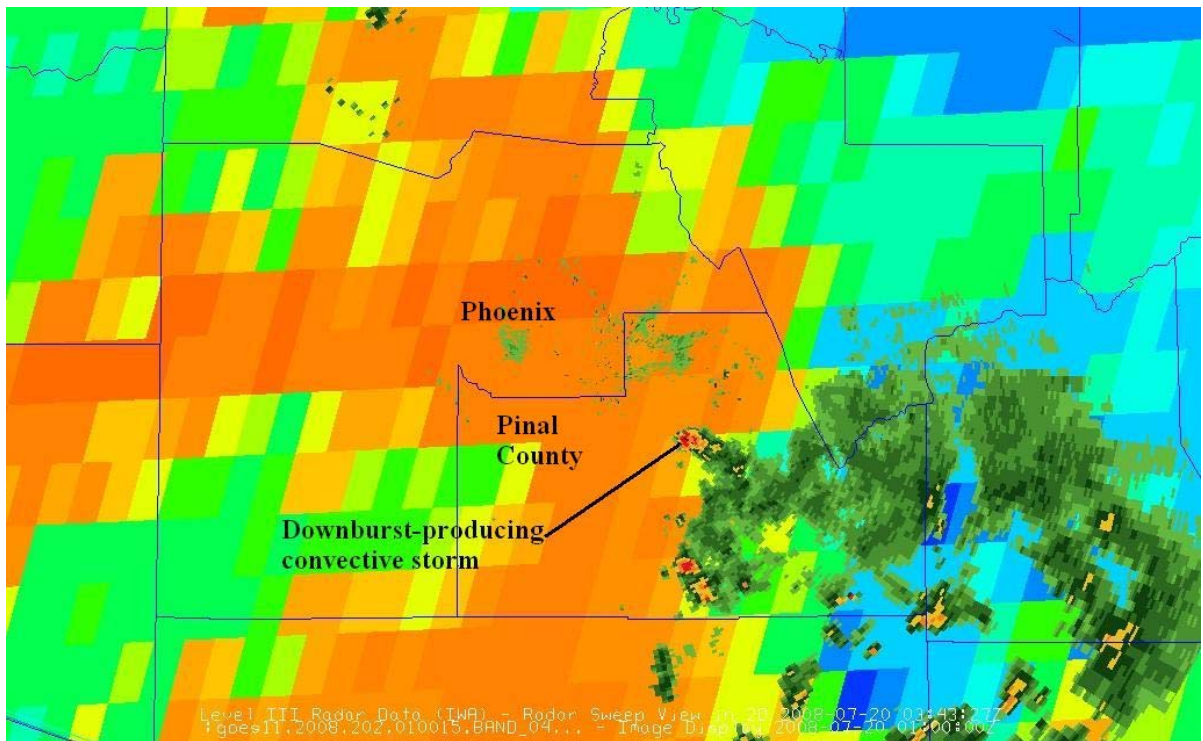

Figure 1. Color-enhanced GOES-11 imager microburst risk product at 0100 20 July 2008.

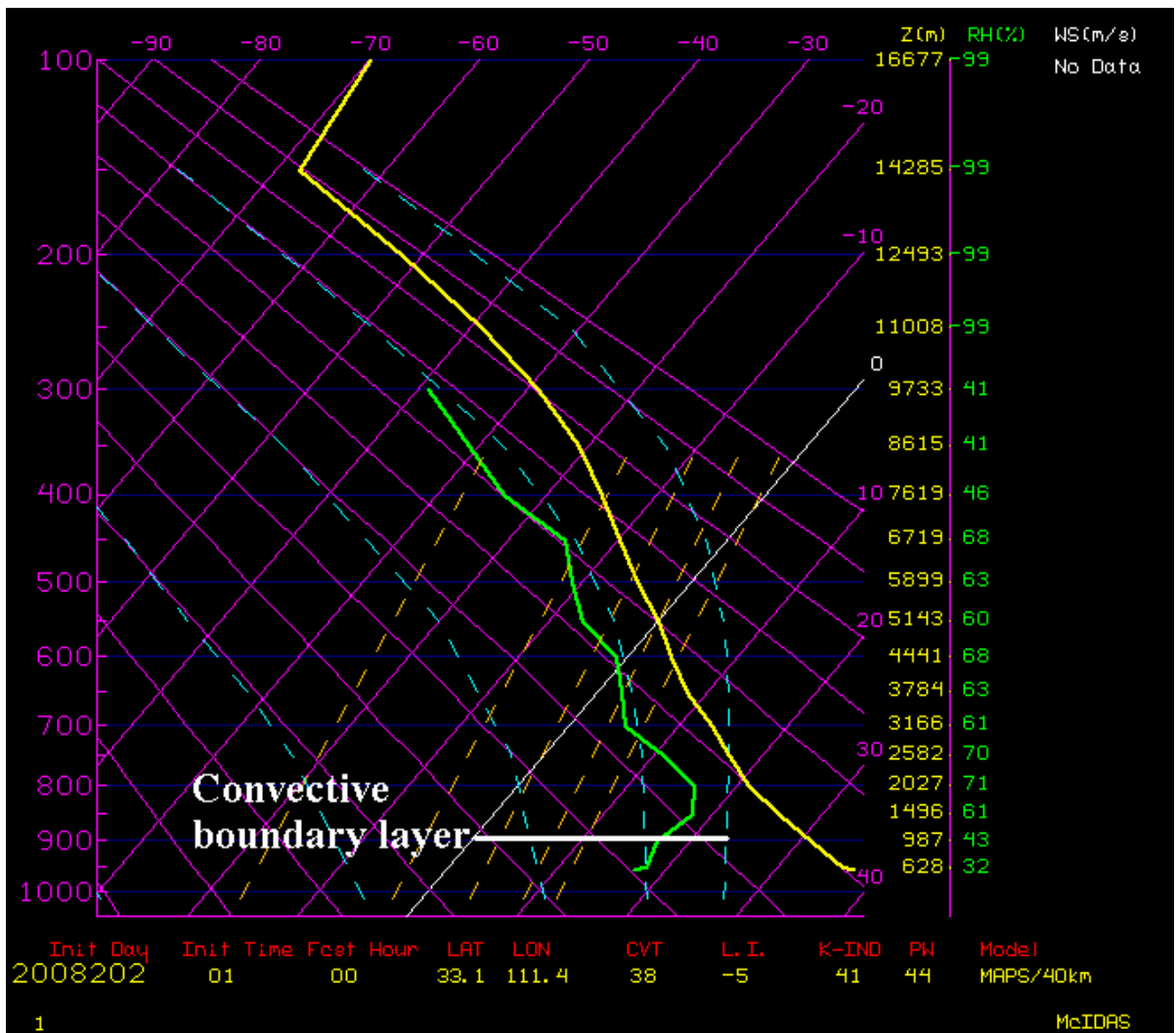
Figure 2.  Vertical sounding profile derived from Rapid Update Cycle (RUC) model analysis data at 0100 UTC 20 July 2008 over Magma FRS ALERT station.

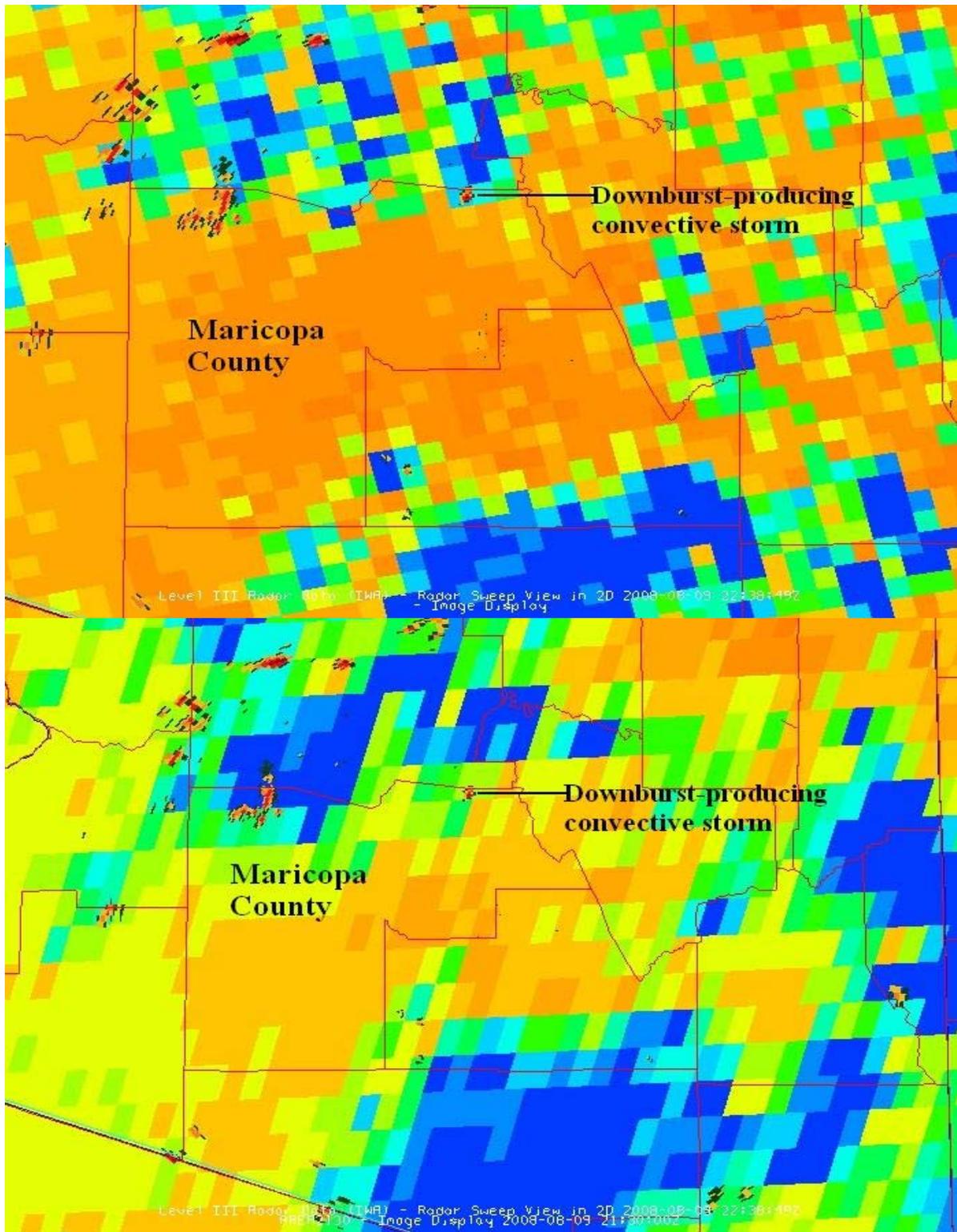

Figure 3. MODIS (top) and GOES (bottom) microburst product images at 2035 UTC and 2130 UTC, respectively, with overlying radar reflectivity from Phoenix, Arizona NEXRAD (IWA) at 2238 UTC

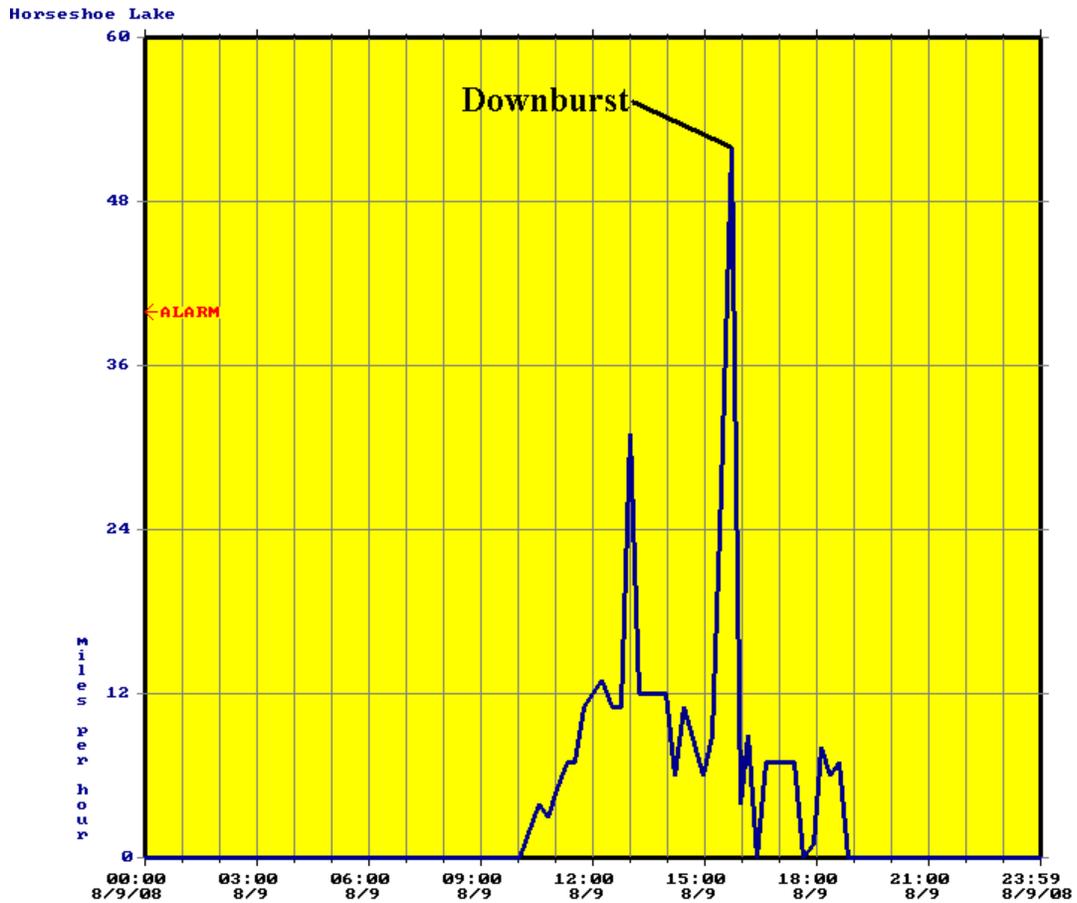

Figure 4. Wind histogram at Horseshoe Lake ALERT station represents downburst occurrence as a sharp peak in wind gust speed near 1540 LST 9 August 2008.

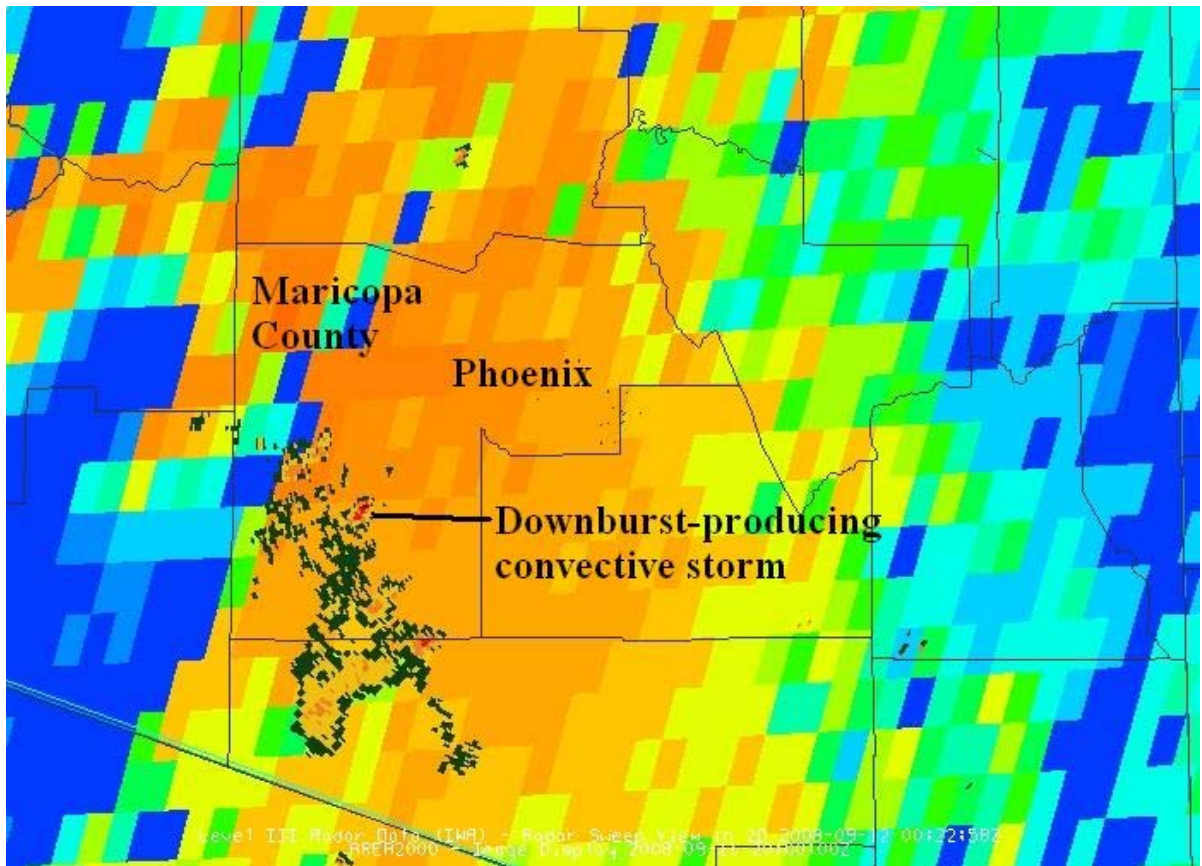
Figure 5. GOES-11 imager derived microburst risk product at 2000 UTC 11 September 2008 with overlying radar reflectivity from Phoenix, Arizona NEXRAD (IWA) at 0022 UTC 12 September.

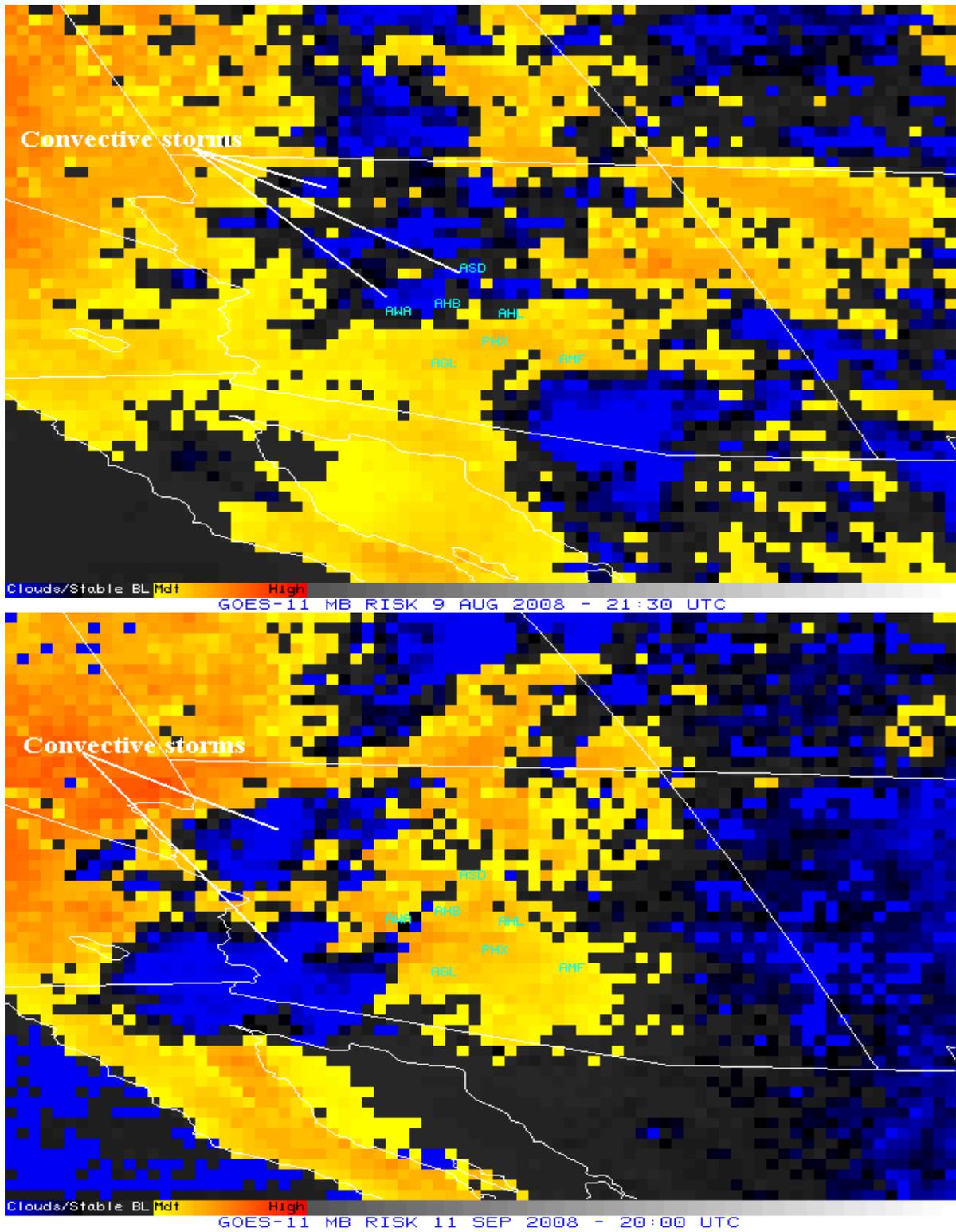

Figure 6. Examples of southwestern U.S. GOES-11 imager microburst risk products that were generated prior to significant downburst events over central and southern Arizona during August and September 2008.